\documentclass[conference]{IEEEtran}    %
\IEEEoverridecommandlockouts

\usepackage[ruled]{algorithm2e}

\usepackage{graphicx}
\usepackage{epstopdf}

\usepackage{subfigure}
\usepackage{amsmath,epsfig}
\usepackage{color}
\usepackage{cite}
\usepackage{algpseudocode}

\newtheorem{observation}{Observation}

\newcommand{\rev}[1]{{\color{black}#1}}

\begin{document}\sloppy

\title{Joint Optimization of Data Sponsoring and Edge Caching for Mobile Video Delivery}
%
\author{
\IEEEauthorblockN{Haitian Pang\IEEEauthorrefmark{1},
Lin Gao\IEEEauthorrefmark{2},
Lifeng Sun\IEEEauthorrefmark{1}}
\IEEEauthorblockA{\IEEEauthorrefmark{1}
Tsinghua National Laboratory for Information Science and Technology, and
\\
Department of Computer Science and Technology, Tsinghua University, Beijing, China}
\IEEEauthorblockA{\IEEEauthorrefmark{2}
School of Electronic and Information Engineering, Harbin Institute of Technology (Shenzhen), Guangdong, China}
\IEEEauthorblockA{Email: \{pht14, sunlf\}@mail.tsinghua.edu.cn, gaolin@hitsz.edu.cn}
\thanks{The work of H. Pang, and L. Sun is supported by the NSFC under Grant No.61272231 and 61472204, and Beijing Key Laboratory of Networked Multimedia. }

}

\maketitle

\begin{abstract}
In this work, we study the joint optimization of edge caching and data sponsoring for a video content provider  (CP), aiming at reducing the content delivery cost and increasing the CP's revenue.
Specifically, we formulate the joint optimization problem as a two-stage decision problem for the CP. In Stage I, the CP determines the edge caching policy (for a relatively long time period). In Stage II, the CP decides the real-time data sponsoring strategy for each content request within the period.
\rev{We first propose a Lyapunov-based online sponsoring strategy in Stage II, which reaches 90\% of the offline maximum performance (benchmark).
We then solve the edge caching problem in Stage I based on the online sponsoring strategy proposed in Stage II, and show that the optimal caching policy depends on the aggregate user request for each content in each location.} Simulations show that such a joint optimization   can increase the CP's revenue by 30\%$\sim$100\%, comparing with the purely data sponsoring (i.e., without edge caching).

\end{abstract}

\section{Introduction}
\label{sec:intro}

\subsection{Background and Motivations}

Nowadays, we are witnessing the explosive growth of global mobile data traffic, which has reached 2.5 exabytes per month in 2014, where  mobile video traffic accounts for 55\% of the total mobile traffic.
According to  Cisco \cite{cisco}, mobile video traffic will contribute nearly 75\% of worldwide mobile data traffic by 2019.
The fast increase of mobile video traffic creates huge opportunities, and meanwhile brings additional challenges for video content providers (CPs).
\rev{Mobile users are often sensitive to the data consumption. Thus, it is important for CPs to design proper incentive mechanisms in order to attract mobile video users \cite{lin-1,lin-2}.}

\rev{\emph{Data Sponsoring} is a novel and effective method recently introduced by CPs to expend user demand for video contents \cite{sen2013smart, sen2013survey, sdp1,sdp2,sdp3,pricing, andrews2014calculating}.
The key idea is to allow CPs to subsidize video users' cost of downloading video data, hence attract more mobile video users and traffic.
With data sponsoring, mobile users benefit from the free wireless access for video contents, and CPs benefit from the increased video users and  traffic (through, for example, selling more built-in advertisements).
That is, it can achieve a win-win situation for CPs and users, and hence has attracted  the interests of both academy and industry.
As a real-world example, AT\&T announced its sponsored data program in January 2014, in which AT\&T allows advertisers to sponsor mobile data to entice users to watch ads they might otherwise have avoided \cite{att}.

 \begin{figure}[t]
 	\centering
	 \includegraphics[width=0.42\textwidth]{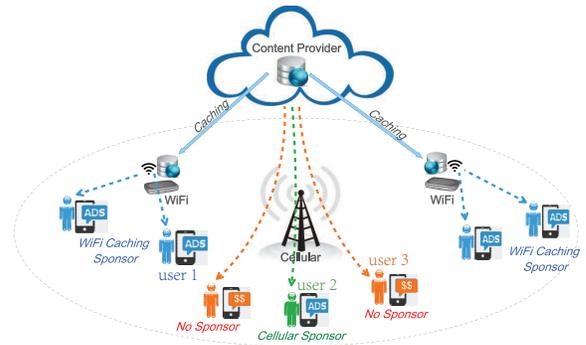}
	 \caption{System Model.}\label{fig:architecture}	
\vspace{-5mm}
\end{figure}
Traditional data sponsoring can potentially help a CP to attract more video traffic and earn more revenue,
but at the same time, it will increase the CP's cost of delivering contents to video users, and also increase the burden of CDNs due to the increased video traffic.
\emph{Edge caching} emerges as a promising paradigm to alleviate the burden of CDNs, reduce the content delivery cost of the CP, and in addition, save the energy consumptions of mobile users \cite{edgecaching,thunder, ec2}.
The key idea is to cache popular video contents on edge networks in advance, and deliver the cached contents on edge networks to the local users through WiFi or Femtocell links  directly.\footnote{Edge caching is different from mobile data offloading \cite{gao-1,gao-2,gao-3}, where CPs deliver \emph{the contents on the server} to mobile users through  local WiFi.}
Edge caching can alleviate the traffic burdens of congested cellular networks and reduce the energy consumptions of mobile users, and hence can deliver contents with a lower cost than the traditional CDNs.
As a real-world example, Xunlei, one of the largest online content download service providers in China, has adopted a new service that  utilizes  users' bandwidths and storage capacities to implement edge caching \cite{thunder}. Xunlei offers the edge network resources to CPs, allowing them to replicate contents for the accesses of neighbour users, and the CPs pay Xunlei fee in return.}

Although there have been a lot of works studying sponsored data \cite{sdp1,sdp2,sdp3,pricing, andrews2014calculating} and the edge caching \cite{edgecaching,thunder, ec2},  all of these works, as far as we know, considered sponsored data and the edge caching separately.
\rev{In this work, we aim to solve the CP payoff maximization problem by jointly considering sponsored data and  edge caching   under the fixed budget \cite{sdp2}. }

\subsection{Solution and Contributions}

We consider a simple yet representative model, where one CP offers video services to a set of mobile users, using edge caching and sponsored data. Each user moves and chooses a video content randomly in a particular time. As illustrated in Fig.~\ref{fig:architecture}, a content can be delivered to a user in two different ways: cellular network and WiFi network (if the content is cached on the local WiFi network in advance).

In the former case, the CP can further decide whether to sponsor the content request via the cellular network. If so, the CP pays the data transmission cost in the cellular network for the user (e.g., user 2 in Fig.~\ref{fig:architecture}); otherwise, the user pays the data transmission cost by himself as usual (e.g., user 3). In the latter case, it is implied that the content is sponsored, as the user does not need to pay for the data transmission cost in the WiFi network (e.g., user 1). Moreover, if a content is sponsored (either via cellular or WiFi), it will be delivered to the user with certain \emph{value-added service} (e.g., advertisement), hence can bring additional value for the CP.

Thus, the key problems for the CP are following:
\subsubsection{WiFi Caching Problem}
Whether to cache a video content on a third-party WiFi network at a particular location for a long time period (e.g., one day)?
\subsubsection{Real-time Sponsoring Problem}
Whether, and if so, how (via cellular or WiFi) to sponsor a video content at each instantaneous time slot (e.g., one minute)?

In this work, we will study the joint optimization of sponsored data and edge caching for maximizing the CP's revenue systematically.
More specifically, we formulate the problem as a two-stage decision problem for the CP.
In Stage I, the CP determines the WiFi caching policy at the beginning of each time period.
In Stage II, the CP determines the real-time sponsoring solution for each content request in each time slot.
As far as we know, this is the first work that studies such a joint optimization of sponsored data and edge caching.
The key contributions of this works are summarized as follows.
\rev{
\begin{itemize}

  \item \emph{Novel Model}: To our best knowledge, we are the first to study the joint optimization of data sponsoring and edge caching for video content providers.
  \item \emph{Offline Optimal Performance (Benchmark):} We formulate the joint optimization problem of edge caching and data sponsoring. The solution shows that the CP's revenue can be increased by $30\% \sim 100\%$ comparing with that without edge caching. Moreover, data sponsoring can support $30\%$ of overall user requests.
  \item \emph{Online Suboptimal Performance:} We propose an online caching strategy and a Lyapunov-based online algorithm to achieve the real-time sponsoring decisions, which outperforms baselines $10\% \sim 50\%$ and reaches $90\%$ of the offline maximum benchmark performance.
\end{itemize}}

The rest of this paper is organized as follows. \rev{
In Section 2, we present the system model.
In Section 3, we provide the problem formulation and the offline solution.
In Section 4, we provide the detailed   online solution. }
We provide the simulation results in Section 5, and conclude in Section 6.

\section{System Model}

\subsection{Network Model}

We consider a mobile video content delivery network (CDN), where a video content provider (CP) provides video service to a set $\mathcal{U} = \{1,2,...,U\}$ of mobile users.
Each user moves   and requests video contents randomly according to his instantaneous individual preference.
Let $\mathcal{S} = \{1,2,...,S\}$ denote the set of all video contents.
A video content can be delivered to a user in two different ways:

\rev{\begin{itemize}
  \item \emph{Cellular Direct Delivery}: The video content (located in the remote  server) will be delivered to the user through the cellular link directly;
  \item \emph{WiFi Cache Delivery}: The video content is cached in a local WiFi network in advance, and will be delivered to the (nearby) user through the WiFi link locally.
\end{itemize}}

For clarity, we illustrate such a dual-channel CDN in Fig.~\ref{fig:architecture}. 
Without loss of generality, we assume that the cellular network covers the whole area, while each WiFi network covers a small area.
We further assume that the coverage areas of different WiFi networks are non-overlap.\footnote{When multiple WiFi networks are overlapping, we can simply choose one of them for video caching.}
Let $\mathcal{L} = \{ 1,  ..., L\}$ denote the set of   areas (locations) covered by WiFi networks.

To enable the WiFi cache delivery, the CP needs to cache the video contents in the corresponding WiFi networks (at the corresponding locations) in advance. \rev{Without loss of generality, we consider the cache operation on a \emph{daily basis} \cite{hulu}. That is, the CP makes the cache decision at the beginning of each day, and each video cache (on a WiFi network) will be available for the whole day. \footnote{Note that there are also some works considering the dynamic cache (e.g., in \cite{dynamiccache}). In this work, we use the fixed cache strategy to emphasize the relationship between caching and sponsoring.}}

To facilitate the user modeling, we further consider a \emph{time-slotted} system, where each daily period is divided into a set $\mathcal{T} = \{ 1,2,...,T\}$ of small time intervals (e.g., one second), called \emph{time slots}.
Each user moves and requests a content randomly in each time slot.
Let $s_u[t] \in \mathcal{S} \bigcup \{0\}$ denote the content request of user $u$ in time slot $t$, where $s_u[t] = 0$ denotes that the user does \emph{not} request any video content.
Let  $l_u[t] \in \mathcal{L} \bigcup \{0\}$ denote the  location of user $u$ in time slot $t$, where $l_u[t] =0$ denotes that the user is in the location without any WiFi coverage.
Then, the service request vector and location vector of user $u \in \mathcal{U}$ in the whole daily period can be written as $\mathcal{S}_u=\{s_u[1], s_u[2], ...s_u[T] \}$ and
$\mathcal{L}_u=\{l_u[1], l_u[2], ...,l_u[T] \}$, respectively.

\subsection{Sponsoring Scheme}

When a user $u$ requests a content $s_u[t]$ at time slot $t$, the CP can decide whether to sponsor the request, and if so, how to sponsor the request.

Moreover, if the CP decides to sponsor a content request, there are two different sponsoring schemes, corresponding to two content delivery schemes, respectively. Namely,

 \begin{itemize}
   \item \emph{Cellular Sponsoring}:
   The CP delivers the content in the remote server to the user through the cellular link, and pays for the cellular delivery cost.
   \item \emph{WiFi Sponsoring}:
   The CP delivers the content cached in the local WiFi network to the user through the WiFi link locally, and pays for the WiFi caching cost.
 \end{itemize}

Let $V_s$ denote the sponsoring value for the CP when sponsoring a video content $s \in \mathcal{S}$, which is related to the value-added service associated with the video. For example, a popular video is usually associated with a high-value advertisement, and hence has a large value. Let $C_s$ denote the cellular delivery cost for the CP when sponsoring a video content $s \in \mathcal{S}$ through the cellular network. \rev{$C_s$ is proportional to the size of the content and the cellular data price.} Let $C_s^w = \alpha \cdot C_s$ denote the WiFi caching cost for the CP when caching a video content $s \in \mathcal{S}$ in a WiFi network. Note that $C_s^w$ is a one-time caching cost, which is independent of the actual times of WiFi sponsoring for content $s$. We define $\alpha$ as the \emph{caching-to-delivery} factor, denoting the relative cost of WiFi caching to cellular delivering. Note that $\alpha > 1$ because caching cost consists of transmission cost as well as storage cost.

\subsection{CP Model}

For each content request, the CP decides whether to sponsor the request, and if so, how to sponsor the request. To do this, the CP need to determine the following strategies:
\begin{itemize}
  \item \emph{Caching Strategy} (at the beginning of each day): Which contents would be cached, and at which locations?
  \item \emph{Sponsoring Strategy} (for each request in each time slot):
      Whether to sponsor each content request, and if so, with which sponsoring scheme?
\end{itemize}

Let $Z[l,s] \in \{0, 1\}$ denote whether to cache a video content $s\in \mathcal{S}$ at location $l\in \mathcal{L}$. Then, given the caching strategy $\{Z[l,s], \forall l\in \mathcal{L}, \forall s\in \mathcal{S} \}$, the total \emph{WiFi caching cost} is
\begin{equation}
\gamma = \sum_{l \in \mathcal{L}} \sum_{s \in \mathcal{S}}  Z[l,s] \cdot  C_s \cdot \alpha.
\end{equation}

Let $y_u[t] \in \{0, 1\}$ and  $x_u[t] \in \{0, 1\}$ denote whether to sponsor the content request (i.e., $s_u[t]$) of user $u$ in time slot $t$  via   WiFi sponsoring   and Cellular sponsoring, respectively.
As only one sponsoring scheme will be chosen, we have:
\begin{equation}\label{eq:c1}
x_u[t] + y_u[t] \leq 1, \quad \forall u\in\mathcal{U} , t \in \mathcal{T}.
\end{equation}
Moreover, the WiFi sponsoring scheme can be chosen only when the content $s_u[t]$ is cached at location $l_u[t]$ (i.e., the location of user $u$ in time slot $t$). Then, we have:
\begin{equation}\label{eq:c2}
y_u[t] \leq Z \big[l_u[t], s_u[t] \big], \quad \forall u\in\mathcal{U} , t \in \mathcal{T}.
\end{equation}
Given the cellular sponsoring strategy  $\{ x_u[t], \forall u\in\mathcal{U}, t\in\mathcal{T}  \}$, the total \emph{cellular delivery cost} in time slot $t$ is:
\begin{equation}
C[t] = \sum_{u \in \mathcal{U}} x_u[t] \cdot C_u[t], \quad \forall t \in \mathcal{T},
\end{equation}
where $C_u[t] = C_{s_u[t]}$ is the cellular delivery cost for sponsoring user $n$'s content $s_u[t]$ in time slot $t$.

Given the cellular and WiFi sponsoring strategies $\{ x_u[t], y_u[t], \forall u\in\mathcal{U}, t\in\mathcal{T}  \}$, the total \emph{sponsoring value} that the CP can achieve in time slot $t$ is:
\begin{equation}
V[t] = \sum_{u \in \mathcal{U}} (x_u[t]+y_u[t] ) \cdot V_u[t], \quad \forall t \in \mathcal{T},
\end{equation}
where $V_u[t] = V_{s_u[t]}$ is the sponsoring value from user $n$'s content $s_u[t]$ in time slot $t$.

Therefore, the CP's overall (time-average) payoff is:
\begin{equation}
R = \frac{1}{T} \cdot \left( \sum_{t=1}^T V[t] -\sum_{t=1}^T C[t] - \gamma \right),
\end{equation}
where the first term is the total (average) sponsoring value, the second term is the total (average) cellular delivery cost, and the last term is the total (average) WiFi caching cost.
For convenience, we denote $R[t] = V[t] - C[t]$ as the instantaneous payoff of the CP in time slot $t$.

\section{Offline Optimization Benchmark}

The CP's objective is to maximize his payoff $R$, given a certain budget. Let $B$ denote the total budget of the CP, for both caching video contents  and  sponsoring video contents. Then, we have the following \emph{budget constraint} for the CP:
\begin{equation}\label{eq:c3}
\sum_{t=1}^T C[t] + \gamma \leq B .
\end{equation}

Therefore, the joint caching and sponsoring problem for the CP can be formulated as follows:
 \begin{equation}\label{eq:maxcom}
 \begin{aligned}
\max \quad & R
\\
s.t. \quad & (\ref{eq:c1}),(\ref{eq:c2}),(\ref{eq:c3});
\\
var.\quad  & x_u[t] \in \{0,1\},\ \forall u, t;
\\
& y_u[t] \in \{0,1\},\ \forall u, t;
\\
& Z[l,s] \in \{0,1\},\ \forall l, s.
\end{aligned}
 \end{equation}

It is notable that (\ref{eq:maxcom}) is an \emph{offline} optimization problem (serving as benchmark), which requires the complete network information\footnote{Note that (\ref{eq:maxcom}) is a binary integer programming, and can be effectively solved by many classic methods.}. Namely, to formulate the problem, the CP needs to know the whole network information in all time slots in advance. In practice, however, the CP cannot obtain the complete network information when making caching and sponsoring decisions.

\emph{We look into the offline optimization problem (\ref{eq:maxcom}) to \rev{gain certain meaningful insight into the design of our two-stage strategy.}} For convenience, we denote $N[l,s]$ as the total number of requests for content $s$ in location $l$, \emph{i.e.,}
$$
N[l,s] = \sum_{u\in \mathcal{U}} \sum_{t\in \mathcal{T}} \mathbf{1}(s_u[t] =s\ \&\ l_u[t] =l),
$$
where $\mathbf{1}(x) = 1$ if $x$ is true, and $0$ otherwise.
From (\ref{eq:maxcom}), we can easily find the following observations.
\begin{observation}
 If $Z[l,s] = 1$, then $x_u[t] = 0$ and $y_u[t] = 1$, for all $  u\in \mathcal{U}, t\in \mathcal{T}$ with  $s_u[t] = s, l_u[t] = l$.
\end{observation}
That is, if a content $s$ is cached in location $l$, then all of the requests for content $s$ at location $l$ will be sponsored by local WiFi. Hence, the CP's payoff achieved from caching a content $s$ in location $l$ \rev{in Stage I} can be computed by:
      $$
      R^w [l, s] = Z[l,s] \cdot ( N[l,s] \cdot V_s - \alpha \cdot C_s );
      $$

\begin{observation}
If $Z[l,s] = 0$, then $y_u[t] = 0$, for all $  u\in \mathcal{U}, t\in \mathcal{T}$ with  $s_u[t] = s, l_u[t] = l$.
\end{observation}
That is, if a content $s$ is not cached in location $l$, then none of the requests for content $s$ at location $l$ will be sponsored via local WiFi.
 Hence, the CP's payoff achieved from sponsoring content $s$ in location $l$ \rev{in Stage II} via cellular is:
      $$
      R^c [l, s] = \sum_{u\in \mathcal{U}} \sum_{t\in \mathcal{T}} \mathbf{1}(s_u[t] =s\ \&\ l_u[t] =l) \cdot x_u[t] \cdot (V_s - C_s)
      $$
      $$
      =  X[l,s] \cdot N[l,s] \cdot (V_s - C_s),
      $$
      where $ X[l,s] = \frac{\sum_{u\in \mathcal{U}} \sum_{t\in \mathcal{T}} \mathbf{1}(s_u[t] =s\ \&\ l_u[t] =l) \cdot x_u[t] }{N[l,s]}  \in [0, 1] $ denotes the percentage of requests for content $s$ at location $l$ that being sponsored via cellular network.

Based on the above, we can transform the original problem (\ref{eq:maxcom}) into an equivalent problem with respect to the decisions regarding each content on each location, \emph{i.e.,} $Z[l,s] \in\{ 0, 1 \}$ and $X[l,s] \in [0, 1 ]$. Formally,
 \begin{equation}\label{eq:maxcom-eq}
 \begin{aligned}
\max \quad  &  R = \sum_{l\in \mathcal{L}} \sum_{s\in \mathcal{S}}
\left( R^w [l, s] + R^c [l, s] \right)
\\
s.t. \quad & X[l,s] + Z[l,s] \leq 1,\ \forall l, s;
\\
&  \gamma + \beta \leq B;
\\
var. \quad  &  X[l,s] \in [0,1],\ \forall l, s;
\\
& Z[l,s] \in \{0,1\},\ \forall l, s;
\end{aligned}
 \end{equation}
where $ \gamma = \sum_{l\in \mathcal{L}} \sum_{s\in \mathcal{S}}  Z[l,s] \cdot \alpha \cdot  C_s
$ is the total WiFi caching cost, and
$ \beta = \sum_{l\in \mathcal{L}} \sum_{s\in \mathcal{S}}
 X[l,s] \cdot N[l,s] \cdot   C_s $ is the total cellular sponsoring cost.
\rev{It is easy to see that problem (\ref{eq:maxcom-eq}) is a mixed-integer linear programming, and we can solve it to get the offline solution.}

\newtheorem{Tm}{Theorem}
\begin{Tm}
Problem (\ref{eq:maxcom-eq}) is equivalent with problem (\ref{eq:maxcom}).
\end{Tm}

\rev{If we solve problem (\ref{eq:maxcom-eq}), we derive $X[l,s]$ and $Z[l,s]$, then we can simply deploy caching strategy by $Z[l,s]$ and randomly choose $X[l,s]*N[l,s]$ user requests in location $l$ for content $s$ to sponsor, \emph{i.e.,} set $x_u[t]=1$. Thus we derive one of the solution of problem (\ref{eq:maxcom}). }
It is important to note that problem (\ref{eq:maxcom-eq}) depends only on the number of requests for each content in each location, i.e., $N[l,s], \forall l\in \mathcal{L}, s\in \mathcal{S} $, while {not} on the request and location of each user in each time slot.
However, it still requires the complete information to compute the exact $N[l,s]$. Nevertheless, it inspires us to find estimations for $N[l,s]$, and   \emph{design the caching policy based on the estimated $N[l,s]$}. \rev{We notice that in problem (\ref{eq:maxcom-eq}) the caching payoff $R^w[l,s]$ and $R^c[l,s]$ are closely coupled with each other, hence maximizing the two parts separately will not derive the optimal overall payoff $R$.}

To this end, we propose a two-stage \emph{online} decision process \rev{based on the analysis in problem (\ref{eq:maxcom-eq})}  for the CP to maximize his payoff without complete network information. In Stage I, the CP determines the WiFi caching strategy at the beginning of the period, based on the \emph{estimated} number of requests for each content in each location. In Stage II, given the WiFi caching strategy in Stage I, the CP determines the sponsoring strategy for each content request in each time slot, using a Lyapunov optimization framework.

\section{Online Optimization Framework}

As mentioned above,  we formulate the CP's payoff maximization problem under incomplete network information as a two-stage \emph{online} decision process:
In Stage I, the CP determines the WiFi caching strategy at the beginning of the period;
In Stage II, the CP determines the real-time sponsoring strategy for each content request in each time slot.
\rev{Next, we analyze these two stages   by backward induction.}

\subsection{Stage II: Best Sponsoring Strategy}

\rev{Lyapunov optimization \cite{neely, gao2015providing} is a widely used technique for solving stochastic optimization problems with time average constraints. In our case, it can be used to maximize the CP's payoff with CP budget constraint in an online manner. We propose a Lyapunov optimization based online sponsoring strategy, which does not rely on the complete network information and converges to the offline benchmark with a controllable approximation error bound.}

In Stage II, the CP determines the best sponsoring strategy for each content request in each time slot, given the WiFi caching strategy $\{Z^*[l,s], \forall l\in \mathcal{L}, \forall s\in \mathcal{S} \}$ determined in Stage I. Hence, the total budget (cost) for WiFi caching is:
$$
\gamma^* = \sum_{l \in \mathcal{L}} \sum_{s \in \mathcal{S}}  Z^*[l,s] \cdot  C_s \cdot \alpha.
$$
And the reminder budget for real-time data sponsoring is:
$$
\widetilde{B} = B -\gamma^* .
$$

The key idea of Lyapunov optimization is to use the \textit{stability} of the queue to ensure that the time average constraint is satisfied. Following this idea, we first introduce virtual queue for the CP budget. Let $q^{t}$ denote the queue backlog of the CP at time $t$. Then, the queue dynamics of CP is:

  \begin{equation}
  q^{t+1} = \Big[q^t - \frac{\widetilde{B}}{T}\Big]^++C[t].
  \end{equation}
By queue stability theorem \cite{neely}, the CP budget constraint is equivalent with $q$'s stability.

  We study the queue stability by using the \emph{Lyapunov drift}, which is defined as follows:
 \begin{equation}\triangle[t] = \frac{1}{2}(q^{t+1})^2 - \frac{1}{2}(q^t)^2.\end{equation}	
By the Lyapunov drift theorem (Th. 4.1 in \cite{neely}), if an algorithm greedily minimizes the Lyapunov drift $\triangle[t]$ in each slot $t$, it potentially maintains the stability of the queue (i.e., ensures the budget constraint of  the CP).

 Next, we analyze the joint queue stability and payoff maximization. By the Lyapunov optimization theorem (Th. 4.2 in \cite{neely}), to stabilize the queues while optimizing the objective, we can use such an allocation policy that greedily minimizes the following \emph{drift-plus-penalty}:
 \begin{equation}
 \Pi[t]= \triangle[t] - \phi \cdot R[t],
 \end{equation}
where  $\phi \ge 0$ is a non-negative control parameter that achieves tradeoff between optimality and queue backlog.
Directly minimizing $\Pi[t]$  is difficult due to the none-linearity of $\triangle[t]$ with respect to the queue backlog.
Hence, we solve an upper bound of $\triangle[t]$, denoted by $\hat{\triangle}[t]$, which is linear and given by:

\begin{equation}
			\hat{\triangle}[t] = \Theta + q^t \cdot \Big(C[t]-\frac{\widetilde{B}}{T} \Big),
\end{equation}
where $\Theta$ is a constant. 
Thus, we can derive the desired online policy by minimizing the upper bound $ \hat{\triangle}[t] - \phi \cdot R[t] $ of drift-plus-penalty in each time slot. Formally, we present the Lyapunov-based Online Policy in Algorithm \ref{algo:1}.

    \begin{algorithm} \label{algo:1}
        \renewcommand{\algorithmicrequire}{\textbf{Initialization:}}
        \caption{Lyapunov-based Online Policy}
        \begin{algorithmic}[1] 
            \Require $\textbf{q}^1=\textbf{0}$

              \For{$i = 1 \to T $}{
                     Allocation Rule:

                     $\quad min \quad \hat{\triangle}[t] - \phi \cdot R[t]$

                     $\quad s.t. \quad 0 \le x_u[t] + y_u[t] \le 1 \quad \forall u \in \mathcal{U}$

                     $\quad \quad\quad \quad Z^*[l_u[t],s_u[t]] \ge y_u[t] \quad \forall u \in \mathcal{U}$

                     $\quad \quad\quad \quad x_u[t],y_u[t] \in \{0,1\} \quad \forall u \in \mathcal{U}$

                     Update Rule:

                     $\quad q^{t+1} = [q^t - \frac{\widetilde{B}}{T}]^++C[t]$}
        \end{algorithmic}
    \end{algorithm}

\subsubsection{Performance Analysis}
We denote the offline optimal payoff (benchmark) as  $R^*$. By Th. 4.1 in \cite{neely}, together with Theorem 1, we can show that the Lyapunov-based Online Policy in Algorithm \ref{algo:1} converges to $R^*$ with a controllable approximation error bound $\mathcal{O}(1/\phi)$. Formally,

\begin{Tm} Let $R[t]$ denote the CP revenue achieved in each time slot $t$ by using  Algorithm \ref{algo:1}. Then,
\begin{equation}
\lim_{T\rightarrow \infty} \frac{1}{T}\sum_{t \in \mathcal{T}}\textbf{E}(R[t]) \ge R^* - \frac{\Theta}{\phi}.
\end{equation}
\end{Tm}

\subsection{Stage I: Edge Caching Strategy}

\rev{The caching strategy in Stage I is dependent on the sponsoring strategy in Stage II, but we cannot obtain the close form of the optimal online sponsoring solution in Stage II. To this end, we aim to derive the caching strategy from problem (\ref{eq:maxcom-eq}). The key idea is to solve problem (\ref{eq:maxcom-eq}) by estimating $N[l,s]$. Next, we provide two estimation methods for $N[l,s]$ under two different system scenarios.}

\subsubsection{Independent and Identically Distributed System}

\rev{In this scenario, each user's locations and content requests are independent and identically distributed (i.i.d.) across different time slots.
Let $\mu_{u,s}$ denote the probability of user $u$ requesting content $s$, and $\eta_{u,l}$ denote the probability of user $u$ moving to location $l$.
Then, we can compute the \emph{expected} number of requests for each content $s$ on each location $l$:
\begin{equation}
\widetilde{N}[l,s] =\sum_{u\in \mathcal{U}}  \sum_{l\in \mathcal{L}} \sum_{s\in \mathcal{S}}  \mu_{u,s} \cdot \eta_{u,l}.
\end{equation}
Substituting the above expected number $\widetilde{N}[l,s]$ into (\ref{eq:maxcom-eq}), we can derive the caching strategy in this scenario.}

\subsubsection{Discrete-Time Markov System}

In this scenario, each user behaviors rather irregularly, but the crowd behavior shows a strong positive correlation among two successive days.
Intuitively, a hot video today is likely to be hot tomorrow.
Let ${N}^{\dag}[l,s]$ denote the number (of requests for content $s$ at location $l$) counted in the previous time period.
Let $P(M|N)$ denote the one-step transition probability of the request number of a content at a location.
Then, we can compute the \emph{expected} request  number for each content $s$ on each location~$l$:
\begin{equation}
\widehat{N}[l,s] =\sum_{M = 0}^{\infty}   P(M|{N}^{\dag}[l,s]) \cdot M .
\end{equation}
Substituting the above expected number $\widetilde{N}[l,s]$ into (\ref{eq:maxcom-eq}), we can derive the caching strategy in this scenario. We analyze a dataset of 3 million watching sessions provided by iQiYi \cite{iqiyi} and observe that the popularity of each content shows strong correlation in time periods, which enables us to predict the future request number in the Markov framework.

\section{Simulation Results}

\subsection{Dataset Description}

\subsubsection{Mobile Video Streaming and WiFi Station Traces}

We use the mobile video streaming trace collected by iQiYi \cite{iqiyi}, one of the most popular online video providers  in China with about 200,000,000 active mobile users. The dataset contains about 190,000 mobile video users with a total of 3,000,000 watching sessions in Beijing within 2 weeks in May 2015. The number of unique video contents is about 100,000.

We also collected the locations of free WiFi Routes (APs) in Beijing via Tencent Mobile Manager \cite{qq}, a mobile phone app with over 400,000,000 users in China, which helps users discover and connect to free WiFi. The dataset contains the information of over 166,000 WiFi APs in Beijing.

\subsubsection{Mapping of the Two Datasets}

In our experiments, we combine these two datasets to explore the WiFi APs that available to users when they request video contents. We assume the signal range of each WiFi AP is 100m, within which mobile users can connect to the AP. Then we map the location of each user at each content request to the nearest AP if the distance between the user and the AP is less than 100m; otherwise, we map the user to the blank location (i.e., $l=0$) without any WiFi coverage.

\subsection{Performance Analysis}
\begin{figure}[h]
	\centering
	 \includegraphics[width=0.42\textwidth]{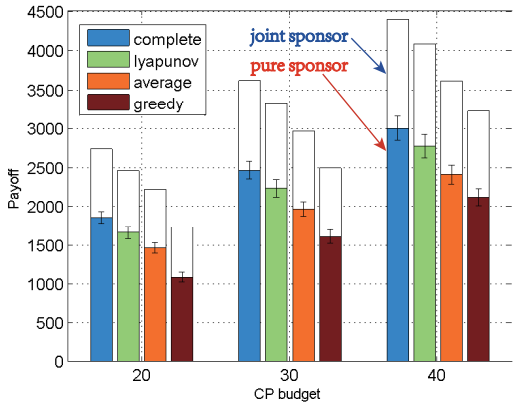}
	 \caption{Pure Sponsoring Payoff under Different CP Budgets}\label{fig:nocache}	
\end{figure}

We compare the payoff under pure sponsor and joint sponsor cases. We notice that joint sponsor outperforms pure sponsor by $30\%\sim 100\%$. We then study the \rev{pure} sponsoring paradigm in Stage II without considering Edge Caching. In order to evaluate the Lyapunov-based algorithm, we choose two sponsoring algorithms as baselines, namely greedy optimization and average greedy optimization. The greedy optimization aims to achieve the best performance in current time slot. The average greedy optimization aims to achieve the best performance and keep the cost even in each time slot. In Fig.~\ref{fig:nocache}, we find Lyapunov-based sponsoring algorithm outperforms baselines in different CP budgets with $10\%\sim 50\%$ gain in CP payoff and can reach about $90\%$ of payoff under complete information. \rev{Average greedy algorithm outperforms the greedy one, because it will distribute the budget in each time slot and tends to sponsor the most valuable contents. Moreover, when CP budget is $40$, the sponsor can cover $30\%$ of users' requests.}

\begin{figure}[h]
	\centering
	 \includegraphics[width=0.5\textwidth]{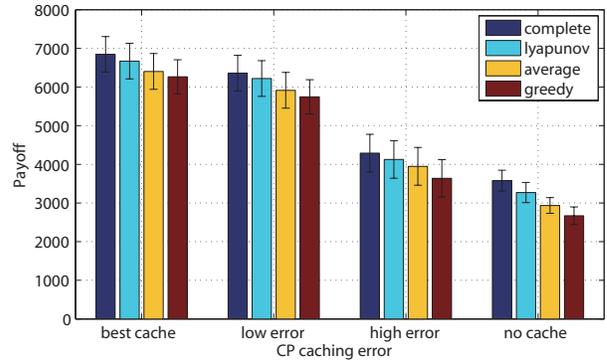}
	 \caption{Joint Payoff under Cache Paradigms}\label{fig:joint}	
\end{figure}

We further study Edge Caching and Sponsored Data jointly. \rev{We achieve the payoff under cache with ``low error'' and ``high error'', which means different prediction accuracies of video broadcast number. In order to evaluate the joint optimization algorithm, we choose two cases as benchmarks:  in ``best cache'', the caching policy is determined under complete information; in ``no cache'', we only deploy pure sponsoring paradigm.} We notice that in the caching cases, sponsoring payoff is less than it is in the no-caching case, because Edge Caching consumes part of CP's budget. Larger prediction error will cause less caching payoff. We can see from Fig. \ref{fig:joint} that Edge Caching and Sponsored Data jointly improve $30\% \sim 100\%$ in CP's payoff than the no-caching case.

\section{Conclusion}
In this paper, we studied the joint optimization of sponsored data and edge caching for maximizing the CP's revenue.
We formulate the joint optimization problem as a two-stage decision problem and solve the problem with  Lyapunov optimization techniques and predicted edge caching principle. The simulation results on a large-scale trace indicate that our design improves CP's payoff significantly.

\end{document}